\documentclass[preprint,prb,showpacs,preprintnumbers,amsmath,amssymb]{revtex4-1}
\usepackage{dcolumn}
\usepackage{graphicx}
\usepackage{bm}
\usepackage[usenames,dvipsnames]{color}
\usepackage{amssymb}
\begin{document}
\author{Simil Thomas}
\author{S. Ramasesha}
\email[]{ramasesh@sscu.iisc.ernet.in}
\affiliation{ Solid State and Structural Chemistry Unit,
 Indian Institute of Science, Bangalore 560 012, India}  
\author{Karen Hallberg}
\author{Daniel Garcia }
\affiliation{Centro At$\acute{o}$mico Bariloche and Instituto Balseiro, 
Comisi$\acute{o}$n Nacional de Energ$\acute{i}$a At$\acute{o}$mica and CONICET, 8400 Bariloche, Argentina}
\title {Fused Azulenes: Possible Organic Multiferroics}
\begin{abstract}
We present compelling theoretical results showing that fused azulene molecules are strong candidates 
for exhibiting room temperature multiferroic behavior, i.e., having both, ferroelectric and ferromagnetic properties. 
If this is experimentally proved, these systems will be the first organic multiferroic materials 
with important potential applications. 
\end{abstract}
\maketitle
\section{Introduction}
Organic electronic materials have come to the fore since the provocative
suggestion of Little~\cite{little-mu,*hiroifrustra,*millerferro}, that a high temperature excitonic superconductor 
can be realized in conjugated molecular systems. 
This has led to the discovery of organic systems which
exhibit a wide range of properties ranging from superconductivity to ferromagnetism.
In recent years, in the field of materials, multiferroics 
have become increasingly important because of the possibility of a wide range of applications.
Most common multiferroics exhibit both, ferroelectric and ferromagnetic (FM) properties.
A few inorganic materials are known to be multiferroics and the most widely known are ferrites and manganites with 
perovskite structures\cite{eerenstein2006,*wang2009}. 
A few purely organic ferroelectrics or exciton multiferroics\cite{Reviews,*TokuraNature} 
and ferromagnets\cite{Sugano} are known.
However, to the best of our knowledge there are no organic multiferroic materials.
In the quest for organic systems that are potential multiferroic materials, 
we have discovered theoretically that fused azulene, a $\rm \pi$-conjugated system, can 
exhibit multiferroic behavior. 

A simple frustrated conjugated molecule that has been well studied is azulene (C$_{10}$H$_8$),
a molecule with fused five and seven-member rings.
It also has a dipole moment $\sim1.0$D attributed to the H\"uckel $``4n+2"$ rule~\cite{wheland}.
Azulene oligomers (n-azulene) can be made up in multiple geometries, 
particularly in fused (n-C$_8$H$_4$) azulene form (see Fig.~\ref{azulene}(a)).
Because of the 5 and 7 member rings of the base monomer all these oligomers are expected to have long-range 
frustration and show a non trivial magnetic state with possible low energy spin excitations.
Besides, depending on the oligomer, the dipole moment of the azulene monomer can also align resulting in a 
molecular ferroelectric phase. 

To explore these molecules, we study the low-lying charge and spin 
energy gaps of azulene oligomers using the long-range interacting Pariser-Parr-Pople (PPP) model and the density 
matrix renormalization group method (DMRG) technique~\cite{dmrgsr,*uli,*karen}.
We show that the existence of a magnetic state in this case does not require the presence of 
flat energy bands, since it also arises  within a spin-1/2 antiferromagnetic Heisenberg 
model (AHM), thus, suggesting  magnetic frustration as the origin.

We find that the fused azulene ground state is a singlet 
for oligomers with up to $5$ azulene units, while with more azulene units the ground state spin increases.
Our studies show that with more than $5$ units and up to $11$ azulene units the oligomers have a 
triplet ground state. 
From these results we predict that the ground state spin of the fused azulene increases with the number of 
azulene units. In the thermodynamic or polymer limit (n$ \to\infty$) we expect fused azulene 
to be a (non-saturated) ferromagnet. We suggest that the ferromagnetism in the model comes 
from the magnetically frustrated geometry of the chain.

In addition, our charge-density calculations show that the ground state of the system has ferroelectric 
alignment of the electric dipoles of the monomeric units and that the charge gap is finite but small 
in the polymer limit of fused azulene. 
Fused azulenes are large but finite molecules and do not have center of inversion 
This leads to the appearance of an electric dipole moment in one direction (sometimes referred to as a 
pyroelectric state). 

These results show that fused azulenes are 
excellent candidates to exhibit multiferroic behavior.
In the next section, we briefly discuss the model Hamiltonian used in our study. 
In the last section we present our results and discussions. 
\section{ Model Hamiltonian for fused azulenes} 
\begin{figure}[ht]
\begin{center}
\includegraphics[height=8.0cm,width=8.5cm]{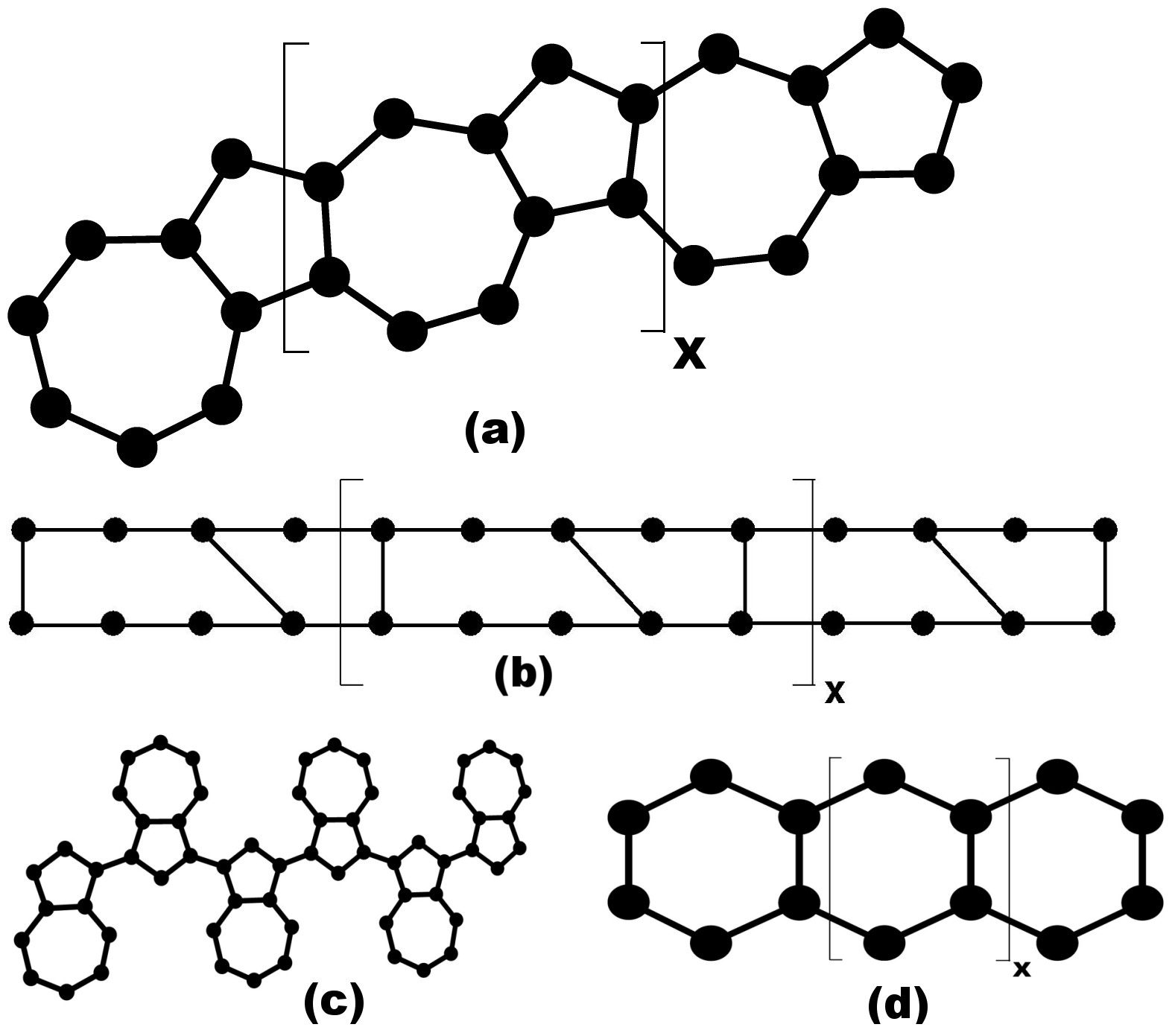} \\ 
\caption{ Structure of (a) fused azulene (b) DMRG ladder-like scheme for building fused azulene 
 (c) Polyazulene and (d)oligoacene.}
\label{azulene}
\end{center}
\end{figure}

The $\pi$-conjugated systems in fused azulene (Fig.~\ref{azulene}(a)), azulene polymer (Fig.~\ref{azulene}(c)) and  
oligoacene (Fig.~\ref{azulene}(d))  
are modeled by considering one $p_z$ orbital on each carbon atom. The PPP  
model~\cite{ppp1,*ppp2} Hamiltonian is given by,
\begin{eqnarray}
\nonumber
H_{PPP}&=&\sum_{<i,j>} t(\hat a_{i,\sigma}^\dagger \hat a_{j,\sigma}^{} ~+~ H.c.) +
  \frac{U}{2} \sum_i\hat n_i(\hat n_i -1 ) \\  
&+& \sum_{i>j} V_{ij}(\hat n_i -1) (\hat n_j-1)
\end{eqnarray}
where $\hat a_{i,\sigma}^\dagger$ ($\hat a_{i,\sigma}$) creates (destroys) an electron on site $i$ with 
spin $\sigma$ and $\hat n_i$ is the number operator and $<i,j>$ implies summation over 
nearest neighbors.

For the fused azulene case, the pentagons and the heptagons are treated as regular polygons of side 1.397 \AA.  
The transfer integral between bonded sites is taken to be -2.4 eV. 
The Hubbard parameter $U$ for Carbon is fixed at 11.26 eV. 
The inter-site interactions are parameterized using the Ohno~\cite{pppohno} formula.
The PPP model with these standard parameters has been  extensively used for successfully modeling the 
excitations in a host of conjugated molecules. This model is used throughout the paper unless otherwise stated.

The relatively large value of $U/t$ allows us to make the approximation of the active phase space by 
 a single electron per site. 
This allows checking the magnetic excitations by using a spin-1/2 AHM.
\begin{equation}
H_{H}=J\sum_{<i,j>} \vec{S}_i\cdot \vec{S}_j ~~~~~ (J>0)
\end{equation}
We study this model to compare magnetic excitations in oligoacenes and fused azulenes.

The DMRG~\cite{dmrgsr,*uli,*karen} method has proved to be ideally suited to study conjugated polymeric systems, 
within model Hamiltonian approaches. 
It has been shown that the DMRG method is accurate even in the presence of long-range interactions 
if these interactions are diagonal in real space, as with the PPP model~\cite{sahoo}. 
Many of the interesting conjugated polymers consist of ring systems, such as phenyl, pyridine, 
thiophene, furan or pyrrole rings. 
Building the desired oligomers of such ring systems by adding two sites at each infinite DMRG step is non-trivial and the 
general method was first illustrated for the case of poly-para-phenylenes~\cite{ppvanu}. 
A similar approach has been used for building the molecular systems in this study.
We have checked our finite DMRG calculations against the non-interacting model results for all oligomers and for 
different cut-offs in the number of dominant density matrix eigenvectors, $m$.
Based on these studies, we find $m=500$ is optimal.
\section{ Results and Discussion}
We have carried out the DMRG calculations on the structures shown in Fig.~\ref{azulene}(a) (fused azulene), 
(using a ladder-like building block (Fig.~\ref{azulene}(b)), Fig.~\ref{azulene}(c) (polyazulene) 
and Fig.~\ref{azulene}(d) (oligoacene). 
For fused azulene the energy per azulene unit linearly extrapolates 
to $-19.117$~eV in the thermodynamic limit.

The oligomers we consider have an even number of sites and a half-filled band, 
implying  states with an integer total spin $S$.
These systems have SU(2) symmetry so we take advantage of the degeneracy of the different spin projections $M_S$ 
for a total spin $S$ to determine the total spin of the system in its ground state. 
 Using the DMRG method it is straightforward to calculate  energy of states with fixed $M_S$.
The ground state spin of the system is {\em S} if the lowest energy states in the $M_S$=0,1,..S
subspaces are degenerate and a gap exists from these to the $M_S$=S+1 state.

\begin{figure}[ht] 
\begin{center}
\includegraphics[width=9cm,height=6cm]{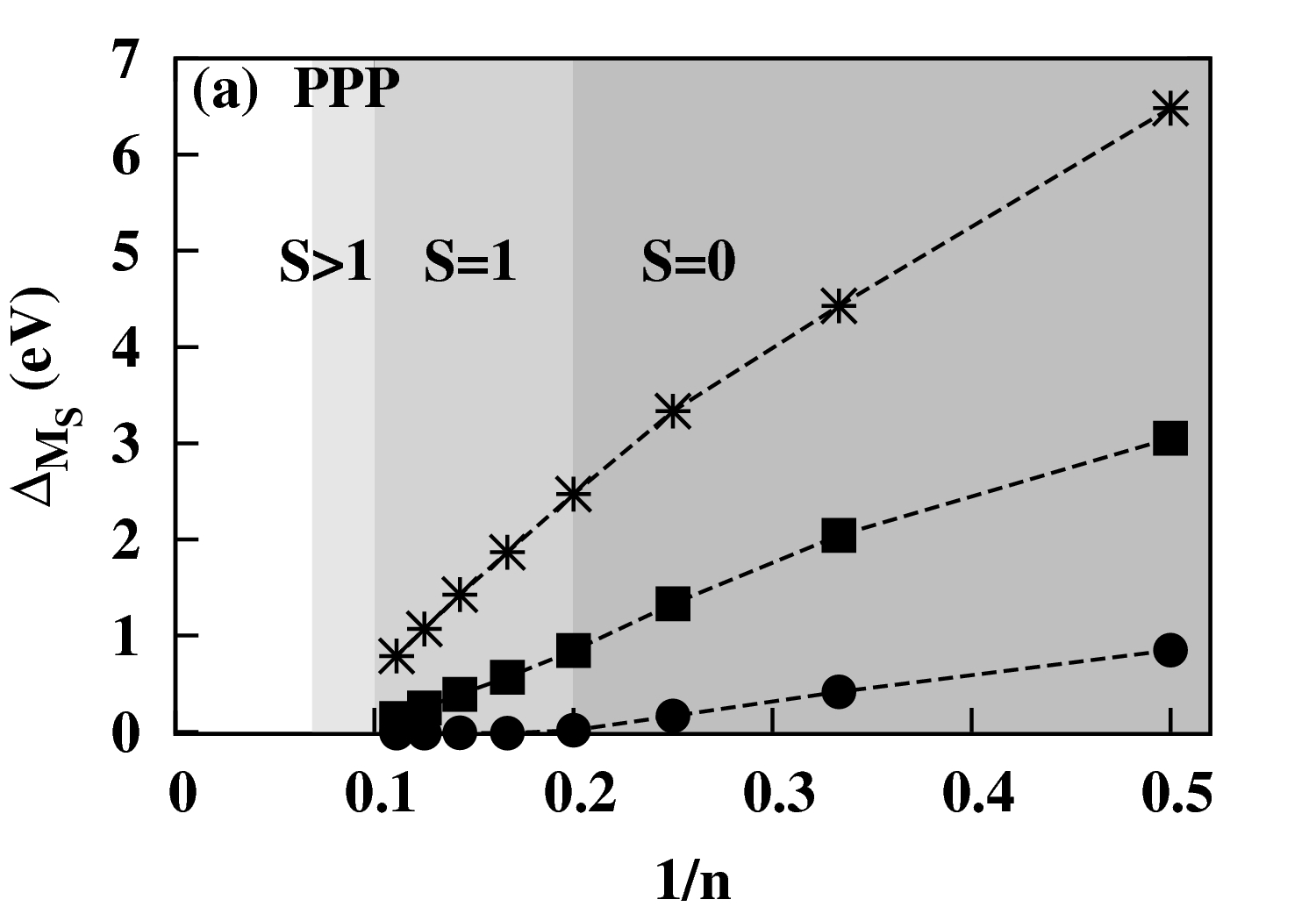}
\includegraphics[width=9cm,height=6cm]{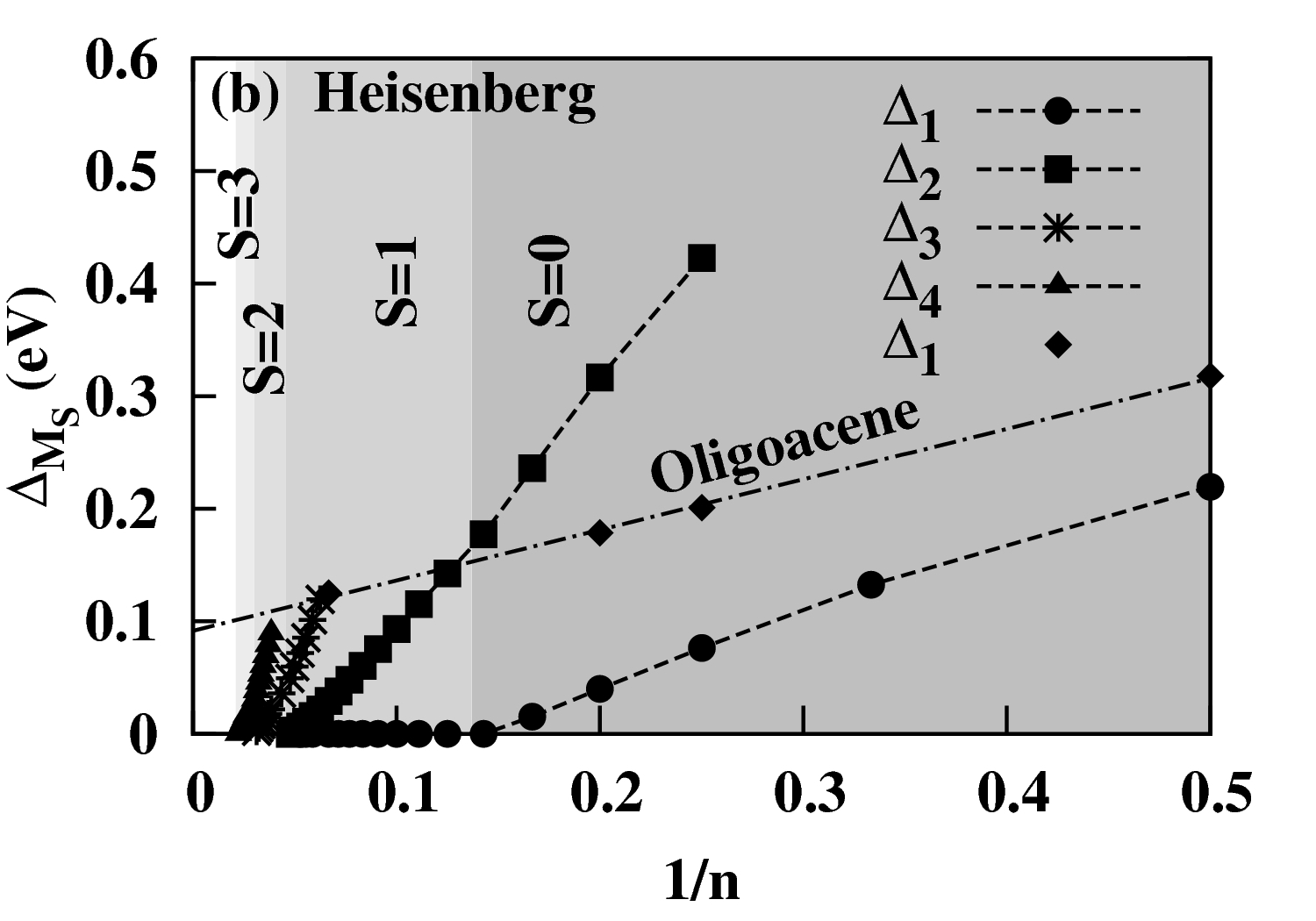}
\caption{Spin gaps (in eV) vs $\frac{1}{n}$ defined as $\Delta_{M_S}=E(M_S)-E(0)$, $M_S=1,2$ or $3$ 
for the PPP model (a) and for the Heisenberg approximation (b). 
The shaded regions represent ground state of a particular spin for the chain lengths above  
which they occur. The spins of states in the shaded region progressively increase in
units of 1 from right to left, for fused azulenes. In case of oligoacenes, $\Delta_1$ 
is always positive, implying a singlet ground state.}
\label{spingap}
\end{center}
\end{figure}

Within the PPP model we computed the lowest energy $\rm E(M_S)$ 
for spin projections $\rm M_S=0,1,2$ and 3 for different  n in fused azulenes.
In Fig.~\ref{spingap}(a) we show the spin gap defined as the energy difference $\rm \Delta_{M_S}=E(M_S)-E(0)$.
Fused azulene oligomers with less than 5 monomers ($\rm n\le 5$)  have total spin $\rm S=0$. 
For larger  oligomers, 
i.e., $\rm 5 < n \le 11$, the lowest $\rm M_S=1$ state becomes degenerate (within the accuracy of the calculation) 
with the lowest $\rm M_S=0$ state indicating a total spin  $\rm S=1$ for the ground state. 
For  even larger oligomers, we have strong indications that lowest energy states in higher 
$\rm M_S$ sectors will become degenerate showing 
transitions to larger spin values for the ground state. 
For example, from the behavior of the gap between the lowest $\rm M_S=2$  
and  $\rm M_S=0$ level, we see that this gap  will also vanish 
when the system size  increases to 
about 10 or 11 azulene units leading to spin $\rm S=2$ ground state. 
The gap $\Delta_3$ between the lowest $M_S=3$ and $M_S=0$ levels also appears to vanish for even larger systems, 
making $S=3$ the total spin of the  ground state. 
Therefore, it is likely that in the polymer limit, a (non-saturated) ferromagnetic ground state will result. 

Polymers with a triplet ground state are, to the best of our knowledge, not known.
Monkman and coworkers~\cite{monkman} 
predicted that a polymer will have a triplet ground state if the lowest singlet exciton 
energy is below $1.3~eV$. This conclusion was arrived at since triplet exciton 
in their systems were $\sim$~1.3~eV below the singlet exciton. 

To see if the system is a polymeric example of flat-band ferromagnetism, we examined the non-interacting 
one-particle states of the systems of up to 20 azulene units. 
 We find that the (HOMO-k) and (LUMO+k) levels come within an energy gap $<$ 0.1t with 
k=0 for $\rm 5 \le n < 10$, k=1 for $\rm 10 \le n < 15$ and k=2 for $\rm 15 \le n < 20$.
However, the degeneracy is not sufficiently close to expect exchange correlation to 
yield a high spin ground state. 

An alternative mechanism for the existence of ferromagnetism for larger systems is magnetic frustration 
stemming from  the lattice geometry~\cite{diep2004}. 
To illustrate this we considered a simple  spin-1/2 AHM  
for fused azulenes with a  spin-1/2 entity at each site. 
As seen in Fig.~\ref{spingap}(b) 
the existence of high spin  ground states  are further confirmed with larger chains using this simplified model. 
While the number of monomers needed to increase the ground state spin 
is  different compared  to the PPP model, 
both models share the same type of transition.
The fact that the purely  AHM reproduces these results is a clear indication 
that this FM state does not have its origin in Fermi level degeneracies.

To further confirm the geometric nature of this FM state we have computed the spin gap $\Delta_1$ 
using the spin-1/2 AHM on the oligoacene lattice(Fig.~\ref{azulene}(d)). This differs from fused azulene  
in the fact that all the fused rings are hexagons.
The geometry allows for a Ising-like non-frustrated antiferromagnetic  ground state 
(which is impossible in systems with odd membered rings).
In this case, we found that the singlet was always the ground state, with a finite gap 
to the lowest triplet state which remained finite in the thermodynamic limit. 
Similar calculations were earlier performed on oligoacenes using the PPP model~\cite{chgap-raghu}. 
It was found that the spin gap in the PPP model remained finite in the polymer limit. 
To explore if extended frustration is essential for a ferromagnetic phase, we carried out
calculations on polyazulene (Fig.~\ref{azulene}(c)). In this case we found a singlet 
ground state for the largest system with eight azulene units in our study 
and a finite spin gap for the thermodynamic system. 
These results show the complex nature of the origin of the high spin states in these geometrically 
frustrated systems. Further studies are necessary to elucidate the nature of the ferromagnetism in 
these cases.  

It is well known that low-dimensional systems can undergo a  distortion of the Peierls' 
type which could lead to stabilization of the  singlet state relative to the  triplet state~\cite{sinha1993direct}.
So for a real molecule it is important to see if distortions could destroy the magnetic state.
Bond-order of a bond (i,j) in the ground state is defined as~\cite{DetailedBond}
$\frac{1}{2} < \sum_{\sigma}\{ a_{i,\sigma}^\dagger a_{j,\sigma}+ 
                               a_{j,\sigma}^\dagger a_{i,\sigma}\} > $.
Bonds with large (small) bond orders will have a tendency to become 
shorter (longer) in the equilibrium geometry.
We note that in both the singlet and the triplet states, the bond orders for the peripheral bonds are nearly uniform. 
This shows that the system is unlikely to undergo a structural distortion.
We also note that the bond orders for the bonds that connect the upper and lower polyene like chains are the 
smallest, implying that any distortion of the system will only uniformly increase the distance between  
the upper and lower chains, which is unlikely to alter energy level ordering of the low-lying states.  

\begin{figure}
\begin{center}
\includegraphics[height=5.0cm,width=8.5cm]{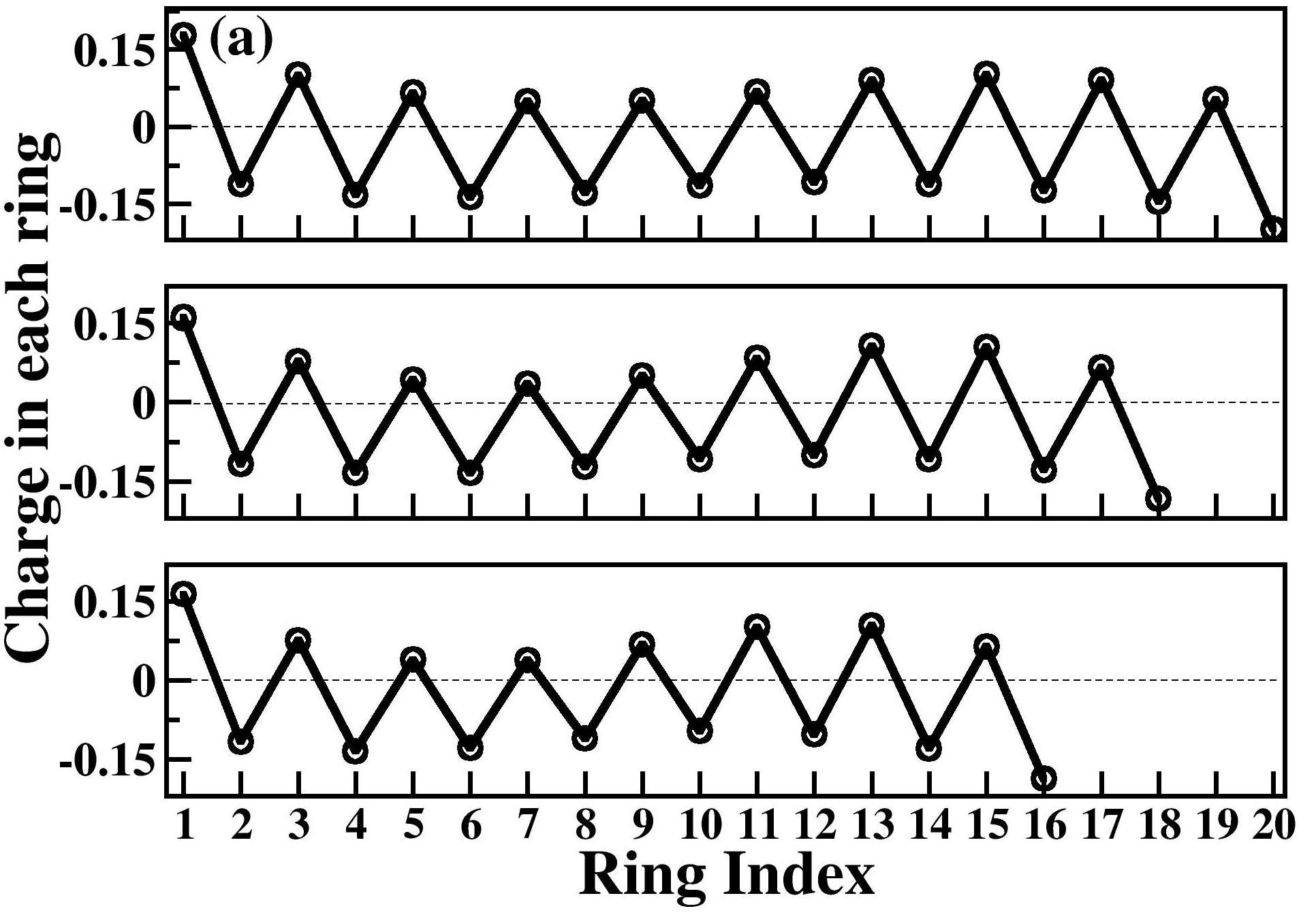}
\includegraphics[height=4.0cm,width=8.5cm]{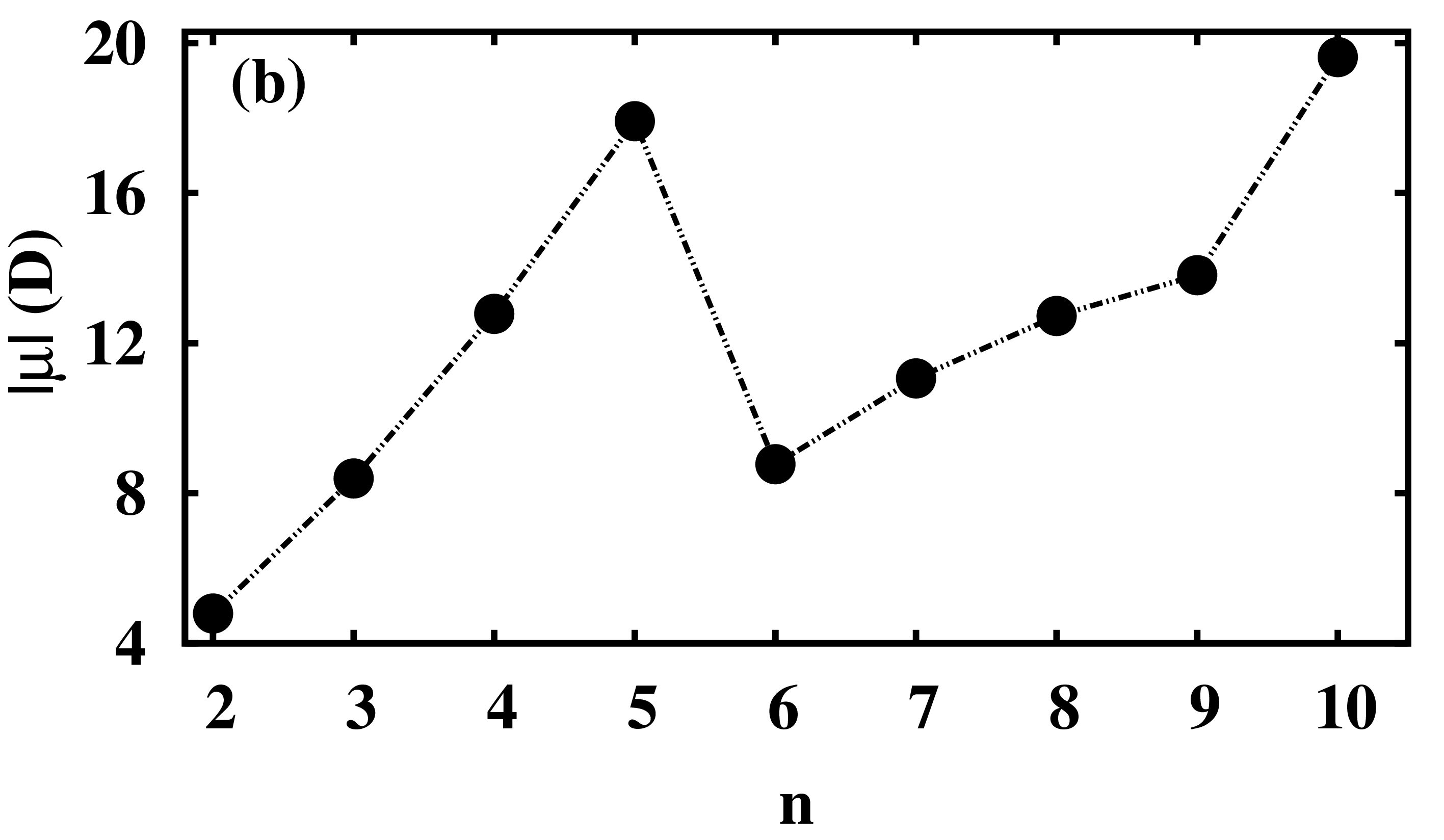}
\caption{(a) Net charge (p-c) in each ring of fused azulenes with n=$10, 9$ and $8$ azulene units 
(top to bottom), see text. 
Even (odd) ring index corresponds to a five (seven) membered ring.
(b) shows dipole moment in the ground state for different number of azulene units.}
\label{chwave}
\end{center}
\end{figure}

To determine the polarized nature of the ground state of fused azulene, we investigate 
the charge distribution and charge gap in these oligomers within the PPP model. 
We define the charge of a ring as (p-c) (p = number of sites in the ring; 
c = total charge in the ring,~~$~\sum_{i~\epsilon~ring}~ <n_i>$) 
for oligomers with up to 10 azulene units.
We show the net charge in the seven- and five-member rings in the oligomers in Fig.~\ref{chwave}(a). 
In all these oligomers, seven member rings have positive charge 
while five member rings have negative charge. 
The amplitude of the charge density wave in the system remains approximately the same, 
independent of the oligomer size, hence we expect a ferroelectric state in the polymer limit.
In Fig.~\ref{chwave}(b) we give the dipole moment ($\vec{\mu}=\sum_i (<n_i>-1)\vec{r}_i$) in the ground state. 
We note that when the spin of the ground state changes, there is a drop in the dipole moment. 
This is due to increase in covalency of the ground state as the spin of the state increases. 
Our results show that the fused azulene system in the polymer limit
also exhibits a spontaneous polarization.  Notice that fused azulene does not have a center of inversion, 
a necessary condition for the emergence of electric dipolar moment. It consists of a sequence of azulene 
monomers containing seven and five-member rings in a definite order. Thus the polymer necessarily 
has a 7-member ring on one end and a 5-member one on the other. 

\begin{figure}
\begin{center}
\includegraphics[height=5.0cm,width=7.0cm]{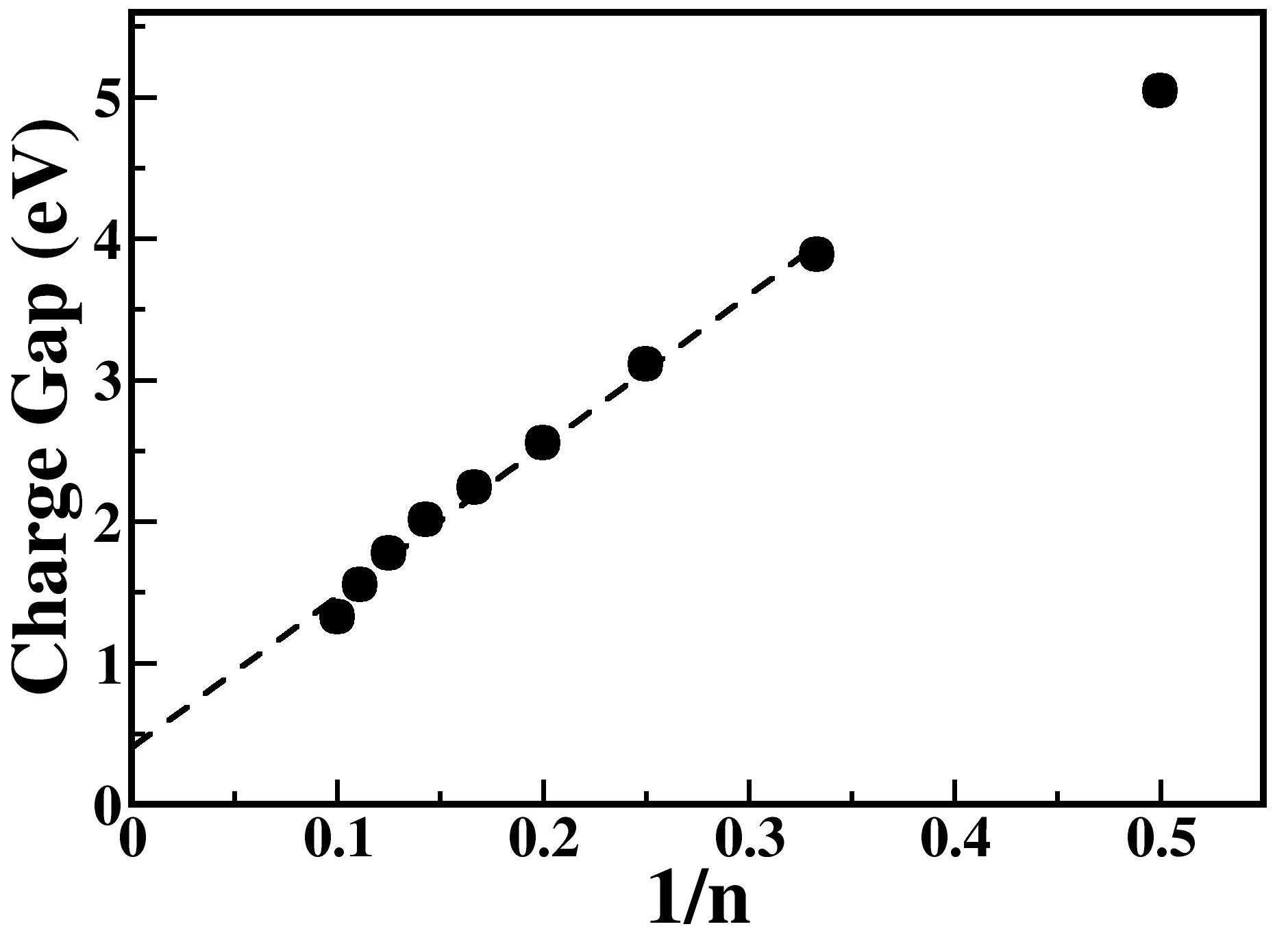}
\caption{Charge gap (in eV) for fused azulenes vs $\frac{1}{n}$.}
\label{chgap}
\end{center}
\end{figure}

The charge gap measures the energy required to create an independent electron-hole pair in a system.
The charge gap for a polymeric system can be obtained by extrapolating $\Delta_c(n)~=~E^+(n)~+~E^-(n)~-~2E^0_{gs}(n)$ 
(where n is the  number of monomers and superscripts `+' and `-' refer to cation and anion while `0' 
refers to the neutral species) to the thermodynamic limit. 
The quantity $\Delta_c(n)$ gives the energy required to create a pair of free moving electron and hole 
in the ground state~\cite{hole}. 
The plot of the charge gap {\it vs} $\frac{1}{n}$ is shown in Fig.~\ref{chgap}.
Extrapolating the charge gap to thermodynamic limit, we obtain the value $\rm \Delta_C(\infty)~=~0.403~eV$. 
This is smaller than typical charge gaps found  for conjugated polymers by almost an order of magnitude~\cite{chgap-raghu}. 
The existence of a charge gap reinforces the conclusion obtained from the ring charge 
disproportionation on the existence of a ferroelectric state in the thermodynamic limit.

To conclude, the results presented in this paper for the ground state of fused azulenes using the 
PPP and Heisenberg models, namely the increasing  ground state magnetic moment with system size, the presence 
of a finite but small charge gap for all lengths and a dipolar moment  
 arising from positively-charged seven-membered and negatively  charged five-membered rings, 
show that this   system is both ferromagnetic and ferroelectric.  Due to the fact that these molecules 
can be large but finite, we expect that real polymers will behave more like single molecular magnets
and electrets. These are compelling results indicating that the fused azulene  system 
could have multiferroic behavior. This would not only signal these 
materials as the first organic multiferroics, but would be also of great 
importance for future organic device applications.

\begin{thebibliography}{10}
\bibitem{DetailedBond}%
  \BibitemOpen
  \bibinfo {journal} {Bond order results will be described in greater detail
  elsewhere.}%
  \bibAnnoteFile{Stop}{DetailedBond}%
\bibitem{hole}%
  \BibitemOpen
\bibfield{journal}{%
    }%
  \bibinfo {journal} {The down turn for large n stems from the error in DMRG
  method, which leads to a reduction in the charge gap due to its variational
  character}%
  \bibAnnoteFile{NoStop}{hole}%
\end{thebibliography}
\end{document}